\documentclass[prl,twocolumn,showpacs,preprintnumbers,amsmath,amssymb,floatfix]{revtex4}

\usepackage{subfigure,graphicx,epsfig,amsmath,amsfonts,amssymb,xcolor,slashed,ulem,multirow}

\newcommand{\be}{\begin{equation}}
\newcommand{\ee}{\end{equation}}
\newcommand{\ba}{\begin{eqnarray}}
\newcommand{\ea}{\end{eqnarray}}
\newcommand{\nn}{\nonumber}
\newcommand{\mev}{\textrm{ MeV}}

\begin{document}

\title{Role of a triangle singularity in the $\pi \Delta$ decay of the $N(1700)(3/2^-)$}
 
\author{L. Roca}
\affiliation{Departamento de F\'isica, Universidad de Murcia, E-30100 Murcia, Spain}

\author{E.~Oset}

\affiliation{Departamento de
F\'{\i}sica Te\'orica and IFIC, Centro Mixto Universidad de
Valencia-CSIC Institutos de Investigaci\'on de Paterna, Aptdo.
22085, 46071 Valencia, Spain}

\date{\today}

\begin{abstract} 
 
We show the important role played by the $\pi\Delta(1232)$ channel in the build up of the $N(1700)(3/2^-)$ resonance due to the non-trivial enhancement produced by a singularity of a triangular loop. The $N(1700)$ is one of the dynamically generated resonances produced by the coupled channel vector-baryon interaction. The $\pi\Delta$ channel was neglected in previous works but we show that it has to be incorporated into the coupled channel formalism due to an enhancement produced by a singularity in the triangular loop with $\rho$, nucleon and $\pi$ as internal loop lines and  $\pi$ and $\Delta$ as external ones.
The enhancement is of non-resonant origin but it contributes to the dynamical generation of the $N(1700)$ resonance due to the non-linear dynamics involved in the coupled channel mechanisms. We obtain an important increase of the total width of the $N(1700)$ resonance when the $\pi\Delta$ channel is included and provide  predictions for the partial widths of the $N(1700)$ decays into $VB$ and  $\pi\Delta$.

\end{abstract}

\maketitle

\section{Introduction}

The $N(1700)$($3/2^-$)  is catalogued in the PDG \cite{pdg} as a three star resonance, in spite of which there is a large dispersion of the results for its properties from different groups as can be seen in Table~\ref{tab:intro}.
\begin{table}[h]
\begin{center}
\begin{tabular}{|c|c|c|c|c|c|}
\hline    Ref.          &   Mass [MeV]    & Width [MeV]     &  \begin{tabular}[c]{c}$\Gamma_{\pi \Delta}/\Gamma$\\  S-wave \end{tabular} 
  & \begin{tabular}[c]{c}$\Gamma_{\pi \Delta}/\Gamma$ \\  D-wave \end{tabular} & \begin{tabular}[c]{c}$\Gamma_{\rho N}/\Gamma$\\  S-wave \end{tabular}  \\ \hline
       \cite{sokho} & $1800\pm 35$    & $ 400\pm 100$     &  50-80\%      	 &  4-14\%		&  	        \\
       \cite{cut}   & $ 1675\pm 25 $  & $ 90\pm 40$     & 			 &			&		        	 \\
       \cite{hoe}   & $ 1731 \pm 15 $ & $ 110 \pm 30 $  &  			 &			&		        \\
       \cite{ani}   & $ 1790\pm 40 $  & $ 390 \pm 140 $ &  $72\pm 23$\%		 &  $<10$\%  	 	&  	        \\
       \cite{shre}  & $ 1665\pm 3 $   & $ 56\pm 8$      &  $31\pm 9$\%		 &  $3\pm 2$\%		&  $38\pm 6$\%	        	\\
       \cite{bat}   & $ 1817\pm 22$   & $ 134\pm 37 $   &  			 &			&		        \\
       \cite{vra}   & $ 1736\pm 33$   & $ 175 \pm 133 $ &  $11\pm 1$\%		 &  $79\pm 56$\%	&  $7\pm 1$\%         	\\  \hline
\end{tabular}
\caption{Mass and width of the $N(1700)$ at the PDG\cite{pdg}. }
\label{tab:intro}
\end{center}
\end{table}
 As one can see, these results are quite different, some times incompatible, and one might be even tempted to think that there could be two states, one around 1700 MeV and the other one around 1800 MeV. The PDG values are M=1650-1750 MeV (1700 average) and $\Gamma=$ 100-250 MeV (150 average).
It also quotes values for the branching ratios chosen from some experiments and analyses. The values quoted for the $\pi \Delta $ branching ratio are those from \cite{sokho} shown in Table~\ref{tab:intro}, and the the $\rho N$ appears as "seen".

     Theoretically there is a nice interpretation for a $J^P=3/2^-$ state around 1700 MeV from the study of the vector-octet baryon interaction in \cite{Oset:2009vf}. Indeed, by using the coupled channels $\rho N,~\omega N,~\phi N,~K^* \Lambda,~ K^* \Sigma$, and the driving force from the local hidden gauge approach \cite{Bando:1984ej,Bando:1987br,Meissner:1987ge}, plus the coupled 
     Bethe-Salpeter equations, which impose unitarity, the vector-baryon (VB) scattering matrix is obtained.  Inspection of the poles shows two structures, one around 1700 MeV, which couples mostly to $\rho N$,  and another one around 1980~MeV, which couples mostly to $K^* \Sigma$. The states appear in S-wave and correspond to $J^P=3/2^-$ states. Although the state obtained at 1700 MeV corresponds formally to a $\rho N$ bound state for the nominal mass of the $\rho$, the consideration of the mass distribution of the $\rho$ in \cite{Oset:2009vf} allows the state obtained to decay into $\rho N$ and the state shows up with a width of around 100 MeV, a bit short of the 150 MeV for the average of the PDG. The other shortcoming of the model is that it does not contain the $\pi \Delta$ channel which experimentally accounts for a large fraction of the width. One should, however, take into account that it is quite common in dynamically generated resonances that some channels relevant in the decay play a  smaller role in the structure of the state. Indeed, one can have a bound state of some component ($\rho N$ in the present case) which decays to some open channel, and the strength of this decay depends much on the phase space available. In the present case, from 1700 MeV there is plenty of phase space for $\pi \Delta$ decay. 

  In the present work we show that the relevance of the $\pi \Delta$ channel is tied to a peculiar mechanism involving a triangle mechanism. The issue of triangle singularities was introduced by Landau \cite{landau} and is catching renewed interest nowadays when a vast amount of empirical information in hadron physics is available, and new examples of triangle singularities are showing up \cite{qzhao}. A triangle diagram appears in the case of a particle A decaying into $1+2$, particle 2 decaying to $B+3$ and particles $1+3$ merging into another particle C. This kind of triangle diagram does not always produce a singularity. It requires that the particles are collinear and that  
the process can occur at the classical level, Coleman-Norton theorem \cite{ncol}. The field theoretical amplitude becomes infinite if the intermediate particles are stable. In real cases, some of these particles have a finite width and the infinity gives rise to a finite peak, which usually has important experimental consequences.
Sometimes the peak resembles pretty much a resonance and can be misinterpreted as such.

  Technically, the amplitudes involving a loop integral are  usually solved using the Feynman parametrization and dispersion relations. An easier and more intuitive formulation has been given recently in  Ref.~\cite{Bayar:2016ftu}, performing analytically the energy integration and looking explicitly to the poles in the remaining integral. A simple equation is obtained to see where a singularity will appear, and the final integral is also easy to perform.

One recent example of a process where the triangle singularities is relevant is the $\eta(1405) \rightarrow \pi\,a_0(980)$ and $\eta(1405) \rightarrow \pi\,f_0(980)$ \cite{wu2,wu1,wu3}. In particular the $\eta(1405) \rightarrow \pi\,f_0(980)$ process is isospin forbidden and it becomes largely enhanced due to a triangle singularity which involves  $\eta(1405) \rightarrow K^*\bar K$ followed by $K^*\rightarrow K\,\pi$ and the  merging of $K\,\bar K$ to give the $f_0(980)$. Another example is the  case  of the ``$a_1(1420)$" originally advocated by the COMPASS collaboration as a new resonance \cite{compass}. Yet, as suggested  in Ref.~\cite{qzhao} and proved in Refs.~\cite{mikha,aceti}, the peak observed at COMPASS comes from the $\pi f_0(980)$ decay of the $a_1(1260)$, via a triangle singularity proceeding through $a_1\rightarrow K^*\bar K$, $K^*\rightarrow K\pi$ and $K\bar K\rightarrow f_0(980)$. Not surprisingly, the $f_1(1420)$ catalogued in the PDG as a resonance, found a similar interpretation as the decay mode of the $f_1(1285)$ into $\pi a_0(980)$ and $K^*\bar K$ \cite{vinicius}. Sometimes a triangle singularity has a direct influence in a reaction even if no resonance has been associated to the peak that it creates. This is the case in the $\gamma p\rightarrow K^+\Lambda(1405)$ reaction \cite{wang}. The process $\gamma p\rightarrow K^*\Sigma$, $K^*\rightarrow K\pi$, $\Sigma\pi\rightarrow \Lambda(1405)$ leads to a peak in the cross section around $\sqrt{s}=2120$ MeV that conventional approaches failed to describe. Another example in this line is the role of the triangle singularity in the $\pi N(1535)$ contribution to the $\gamma p \to \pi^0 \eta p$ reaction close to threshold \cite{elsa}, which in \cite{shuntaro} is described in terms of a triangle singularity involving the $\Delta(1700) \to \eta \Delta$ followed by $\Delta \to \pi N$ and $\eta N \to N(1535)$. 

Similarly, the $f_2$(1810) is also explained as a consequence of the $f_2(1650)\rightarrow K^*\bar K^*$, $K^*\rightarrow \pi K$ and $K\bar K^*$ merging into the $a_1$(1260) \cite{geng}. More examples can be found in Refs.~\cite{hanhart,achasov,Lorenz,adam1,adam2}. 

  In some cases the singularity tied to the decay of a resonance does not lead to a new peak at a different energy, but it comes precisely at the same energy where the resonance appears and reinforces it introducing new decay channels. This is the case of the $N(1875)$ resonance studied in \cite{daris}, which develops a singularity at the same energy from the resonance coupling to $\Sigma^* K$, $\Sigma^* \to \pi \Lambda$ and $K \Lambda \to N(1535)$. In this case the singularity reinforces the resonance adding to it a new relevant channel.  

  There is no way a priori to know if a triangle singularity can be relevant or not in a given process. One must perform the detailed evaluation and see what comes out. In this direction it is worth commenting on the suggestion that the narrow peak of the $J/\psi\,p$ invariant mass at 4450 MeV seen by the LHCb collaboration \cite{penta-ex,chinese} could be due to a triangle singularity with $\Lambda_b\rightarrow \Lambda(1890) \chi_{c1}$, $\Lambda(1890)\rightarrow \bar K p$, $p\chi_{c1}\rightarrow J/\psi p$ \cite{ulfguo,liu}. However, as shown in \cite{Bayar:2016ftu}, assuming the preferred experimental quantum numbers of this peak, $3/2^-$, $5/2^+$, the $\chi_{c1} p\rightarrow J/\psi p$ proceeds with $\chi_{c1} p$ in $p$-wave or $d$-wave and the $\chi_{c1} p$ is at threshold at exactly 4450 MeV, hence, this amplitude vanishes there on shell and the suggested process cannot be responsible for the observed peak.

In the present work we show another example of a triangle singularity appearing at the energy where a resonance already exists. Indeed, the $N(1700)(3/2^-)$ is well accepted and, as mentioned above, can be interpreted as mostly a $\rho N$ bound state. The singularity appears from the mechanism of $ N(1700) \to \rho N$ followed by $\rho \to \pi \pi $ and fusion of $\pi N$ to give the $\Delta(1232)$. The singularity of this process appears around 1700 MeV and introduces the $\pi \Delta$ channel  as an important building block into the wave function of the resonance.

\section{Formalism}

\subsection{The vector-baryon channels in the $N(1700)$}

 The $N(1700)$ resonance, ($I=1/2$, $J^P=3/2^-$), was one among several resonances obtained dynamically in ref.~\cite{Oset:2009vf} from the interaction of the lowest mass vector-meson ($V$) octet  and the lowest mass baryon ($B$) octet in coupled channels. In particular, the $N(1700)$ arises dynamically from the $\rho N$, $\omega N$, $\phi N$, $K^* \Lambda$ and $K^* \Sigma$ interaction with isospin $I=1/2$ and in S-wave. The dominant channel is $\rho N$ and also, but less important, $K^* \Lambda$. It is very important to stress that, using the techniques of the chiral unitary approach,
the $N(1700)$ arises naturally from the coupled channel dynamics without the need to include the $N(1700)$ as an explicit degree of freedom.
We will start from the model of ref.~\cite{Oset:2009vf} and,  in the next section, we will improve upon it by adding the $ \pi\Delta$ channel, which produces the triangular singularity that will contribute significantly to the dynamics that generates the  $N(1700)$. Therefore, for the sake of completeness, we briefly summarize the approach of ref.~\cite{Oset:2009vf} to account for the vector-baryon channel unitarization (we refer to \cite{Oset:2009vf} for further details):

From the $VVV$ and $BBV$ interaction Lagrangians provided by the local hidden gauge symmetry formalism \cite{Bando:1984ej,Bando:1987br,Nagahiro:2008cv}, the $VB\to VB$ tree level potential can be obtained, which, neglecting the momentum transfer versus the baryon mass, is given by
\be
V_{ij}=-\frac{C_{ij}}{4 f^2}(k_i^0+k_j^0) 
\vec \epsilon_i\cdot \vec\epsilon_j
\label{eq:Vij},
\ee
where $i$ ($j$) stands for the initial (final) $VB$ channel, $f=93\mev$ is the pion decay constant, $k_i^0$  the energy of the vector meson of the i-th channel and $C_{ij}$ are coefficients given by

\begin{equation} C_{ij} =\begin{pmatrix}
\ 2  \ & \ 0 \ & \ 0 \  & \ \frac{3}{2} \ & \ -\frac{1}{2} \  \\
   & 0 & 0  & -\frac{\sqrt{3}}{2} & -\frac{\sqrt{3}}{2} \\
   &   & 0  & \sqrt{\frac{3}{2}} &  \sqrt{\frac{3}{2}} \\
   &   &    & 0 & 0 \\
   &   &    &  & 2 
\end{pmatrix} 
\label{eq:Cij},
\end{equation}
with $C_{ij}=C_{ji}$ and the indices are in the order $\rho N$, $\omega N$, $\phi N$, $K^* \Lambda$ and $K^* \Sigma$.

\begin{figure}[tbp]
     \centering
     \includegraphics[width=.99\linewidth]{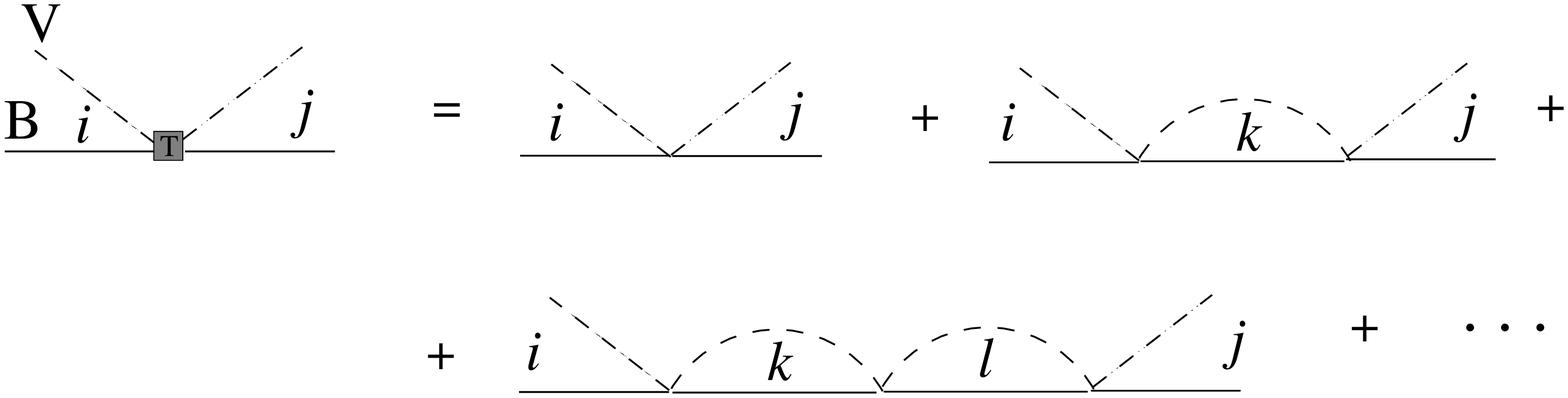}
    \caption{Unitarization of the vector-baryon interaction}
    \label{fig:BS}
\end{figure}

The chiral unitary approach implements exact unitarity in coupled channels to obtain the full  $VB$ scattering matrix using 
the potential of Eq.~(\ref{eq:Vij}) as the kernel of the unitarization procedure. Exact unitarity in coupled channels is implemented by using the Bethe-Salpeter matrix equation (equivalent to the $N/D$ method  \cite{Oller:1998zr,Oller:2000fj}), which has the form
\be
T=[1-VG]^{-1}V
\label{eq:T}
\ee
where $V_{ij}$ is the interaction kernel of Eq.~(\ref{eq:Vij}) and $G$ is a diagonal matrix with $G_i$ the $i$-th vector-baryon loop function, or unitarity bubble.
Equivalently, Eq.~(\ref{eq:T}) represents the resummation shown in Fig.~\ref{fig:BS}.

The vector-baryon loop functions $G_i$ need to be regularized which can be accomplished by means of dimensional regularization in terms of a subtraction constant $a(\mu)$ or with a three-momentum cutoff $q_{\textrm{max}}$. The equivalence between both methods was shown in \cite{Oller:2000fj,Oller:1998hw}.
In ref.~\cite{Oset:2009vf} a subtraction constant $a=-2$ at a regularization scale $\mu=630\mev$ was used, which we have checked 
is equivalent to regularizing with a cutoff of $q_{\textrm{max}}=650\mev$.

In order to account for the finite width of the vector mesons, we fold the loop functions with their corresponding mass distribution provided by the spectral function $\rho_V(s_V)$,
\ba
G_i(s,M_V,M_B) &=& \frac{1}{{\cal N}}
\int_{(M_V-2\Gamma^o_V)^2}^{(M_V+2\Gamma^o_V)^2}
ds_V\,\nn\\
&\times& \rho_V(s_V) \,G(s,\sqrt{s_V},M_B),
\label{eq:convo}
\ea
where $\Gamma^o_V$ is the total width of the vector meson.
In Eq.~(\ref{eq:convo}), ${\cal N}$ is the normalization 
of the spectral distribution:
\be
{\cal N} = 
\int_{(M_V-2\Gamma^o_V)^2}^{(M_V+2\Gamma^o_V)^2}
ds_V\, \,\rho_V(s_V)
\ee
Since the vector meson mass distributions 
are Breit-Wigner like spectra, 
 we take the spectral function 
\be
\rho_V(s_V)=-\frac{1}{\pi}\textrm{Im}
\left\{\frac{1}{s_V-M_V^2+iM_V\Gamma_V(s_V)}\right\},
\label{eq:rho_V}
\ee
where we consider an energy dependent vector meson width
\be
\Gamma_V(s_V)=\Gamma^o_V \frac{M_V^2}{s_V} \left(\frac{q(\sqrt{s_V})}{q(M_V)}\right)^3 \Theta(\sqrt{s_V}-\sqrt{s_V^{th}}),
\ee
with  $q(x)$ the decay momentum of the vector with mass $x$ into the dominant channel, 
and $\sqrt{s_V^{th}}$ the energy of the dominant threshold.

The convolution in Eq.~(\ref{eq:convo}) is crucial to get a finite width for the $N(1700)$ resonance since its mass lies below the lowest $VB$ threshold (in this case $\rho N$) and the width would otherwise be zero. 
Therefore the finite $N(1700)$ width comes essentially from the tail of the $\rho$ meson.

\begin{figure}[tbp]
     \centering
     \includegraphics[width=.95\linewidth]{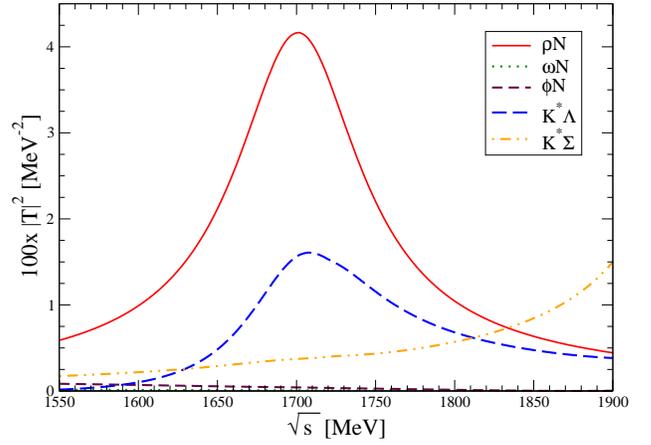}
    \caption{Diagonal vector-baryon amplitudes without the $\pi\Delta$ channel}
      \label{fig:tii}
\end{figure}

In Fig.~\ref{fig:tii} we show the modulus squared of the amplitude of the $VB\to VB$ diagonal  channels in the region around $1700\mev$. We can see the prominent peak of the $N(1700)$ in the $\rho N$ and 
$K^* \Lambda$ channels, which has to do with the couplings to these channels as we will see below. 

\subsection{ The triangle singularity in the $\pi \Delta$ channel}

\begin{figure}[tbp]
     \centering
     \includegraphics[width=.75\linewidth]{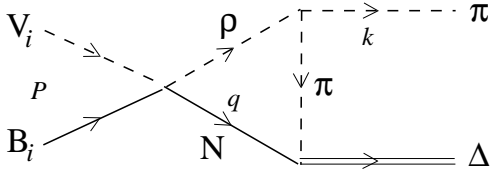}
    \caption{Triangular loop for the $\pi\Delta$ channel.}
    \label{fig:triang}
\end{figure}

As mentioned in the introduction, for particular kinematic conditions \cite{Bayar:2016ftu,ncol}, there can be an important enhancement from a triangular loop for a particular channel which otherwise would be negligible. As we will see, this is the case of 
the $\pi\Delta$ channel in the build up of the $N(1700)$ resonance.
Indeed, let us consider the loop diagram of Fig.~\ref{fig:triang}. 

The value of the incoming energy $\sqrt{s}$ for which the triangular singularity occurs is given by the solution of the following equation \cite{Bayar:2016ftu}
\be
q_{on}-q_{a^-}=0
\label{eq:qomenosqa}
\ee
where $q_{on}$ is the on-shell momentum of the nucleon, $N$, or the $\rho$ meson, in their center of mass frame for a given incoming  $\sqrt{s}$ energy, and 
$q_{a^-}$ 
corresponds to the solution for $\Delta\to N\pi$ in the VB rest frame which has negative imaginary part $(-i\varepsilon)$, with $\vec q$ and $\vec k$ antiparallel.
 (Explicit expressions and further discussion can be found in  \cite{Bayar:2016ftu}).
In the present case the solution of Eq.~(\ref{eq:qomenosqa}) is $\sqrt{s}=1715\mev$, just on the $N(1700)$ mass region. This is why one can expect a priori the $\pi\Delta$ channel to be relevant and then its inclusion in the coupled channels dynamics to generate the $N(1700)$ is called for.

In order to include the $\pi\Delta$ into the coupled channel equation, Eq.~(\ref{eq:T}), we need the $\pi\Delta\to\pi\Delta$ potential, which is given by Eq.~(\ref{eq:Vij}) with $C_{\pi\Delta,\pi\Delta}=5$ \cite{Sarkar:2009kx}, and the non-diagonal $V_{i,\pi\Delta}$ potentials which can be written as
\be
V_{i,\pi\Delta}=V_{i,\rho N}\,V_\textrm{eff}
\label{eq:Vipidelta}
\ee
with $V_\textrm{eff}$ accounting for the
the triangular mechanism in Fig.~\ref{fig:triang} which we evaluate in the following.
For this purpose we need the amplitude for the $\rho\pi\pi$ interaction  and $\pi N\Delta  $ which we obtain from the usual Lagrangians
\ba
{\cal L}_{VPP}&=&-ig\langle[P,\partial_{\mu}P]V^{\mu}\rangle \nn\\
{\cal L}_{\pi N \Delta }& =& - \frac{f_{\pi N \Delta}}{m_\pi}\Psi^\dagger_\Delta S^\dagger_i(
\partial_i\phi^\lambda)T^{\lambda\dagger}\Psi_N  +  \textrm{h.c.}
\ea
with $g=\frac{m_\rho}{2f}$, $f_{\pi N \Delta}=2.13$ and $\vec S^\dagger$ ($\vec T^\dagger$) the 1/2 to 3/2 spin (isospin) transition operators.

Considering the proper isospin $1/2$ combination of the internal $\rho N$ states and of the external $\Delta \pi$ states, the amplitude of the diagram in Fig.~\ref{fig:triang} reads

\ba
-it_i&=& i V_{i,\rho N} \frac{2}{\sqrt{3}} g \frac{f_{\pi N \Delta}}{m_\pi}\nn \\
&&\times 2 M_N \int \frac{d^4q}{(2\pi)^4}
\,\vec\epsilon_i\cdot(2\vec k+\vec q)\,\vec S^\dagger\cdot(\vec k+\vec q)\nn\\
&&\times\frac{1}{q^2-M_N^2+i\varepsilon}\,
\frac{1}{(P-q)^2-m_\rho^2+i\varepsilon}\nn\\
&&\times\frac{1}{(P-q-k)^2-m_\pi^2+i\varepsilon}
\label{eq:tia}
\ea
which is  evaluated in the initial rest frame, $\vec P=0$, 
and $\vec\epsilon_i$ stands for the polarization vector of the vector meson in the initial channel $i$.
The factor $2 M_N$ in Eq.~(\ref{eq:tia}) stems from the use of the Mandl-Shaw  normalization \cite{mandl} for the fermion fields.
After performing the $q^0$ integration, the amplitude in Eq.~(\ref{eq:tia})
takes the form

\begin{align}
-it_i=  V_{i,\rho N} \frac{2}{\sqrt{3}} g \frac{f_{\pi N \Delta}}{m_\pi}
2 M_N
 \left( \vec\epsilon_i\cdot\vec S^\dagger\, t_T^{(1)}
+ \vec\epsilon_i\cdot \vec k \, \vec S^\dagger\cdot\vec k  \, t_T^{(2)}
\right)
\label{eq:tib}
\end{align}
where
\begin{align}
t_T^{(\alpha)}=\int\frac{d^3q}{(2\pi)^3}\frac{H^{(\alpha)}}{8E_N\, \omega_\rho\,\omega_\pi}\,
\frac{1}{k^0-\omega_\pi-\omega_\rho}\,
\frac{1}{P^0-k^0+E_N+\omega_\pi}\,\nn\\
\times\frac{1}{P^0-k^0-E_N-\omega_\pi+i\varepsilon}\,
\frac{1}{P^0-E_N-\omega_\rho+i\varepsilon}\nn\\
\times\left[2P^0 E_N+2k^0 \omega_\pi-2(E_N+\omega_\pi)(E_N+\omega_\pi+\omega_\rho)\right]
\label{eq:ttalpha}
\end{align}
and
\ba
H^{(1)}&=&\frac{1}{2}\left[\vec q\,^2-\frac{(\vec q\cdot\vec k)^2}{\vec k^2}
\right], \nn \\
H^{(2)}&=&2+3\frac{\vec q\cdot\vec k}{\vec k^2}
+\frac{1}{2\vec k^2}\left[3\frac{(\vec q\cdot\vec k)^2}{\vec k^2}-\vec q\,^2\right].
\label{eq:H1H2}
\ea
In Eq.~(\ref{eq:ttalpha}), $\omega_\rho=\sqrt{m_\rho^2+\vec q\,^2}$,
$\omega_\pi=\sqrt{m_\pi^2+(\vec k+\vec q)^2}$ and $E_N=\sqrt{M_N^2+\vec q^2}$.
Except for the functions  $H^{(\alpha)}$, Eq.~(\ref{eq:ttalpha}) is the triangular amplitude $I_1$ of refs.~\cite{Aceti:2015zva,Bayar:2016ftu} but for different masses.

Note that the previous integrals are logarithmically divergent and they  need to be regularized. For consistency with the $VB$ unitarization described in the previous section, which needs regularization of the $VB$ loop for an equivalent cutoff of $q_{\textrm{max}}=650\mev$, we also integrate $|\vec q|$ numerically up to that upper limit. The finite $\rho$ meson width can be taken into account by substituting $\omega_\rho\to\omega_\rho-i\Gamma_\rho/2$  in the propagators of Eq.~(\ref{eq:ttalpha}).

\begin{figure}[tbp]
     \centering
     \includegraphics[width=.95\linewidth]{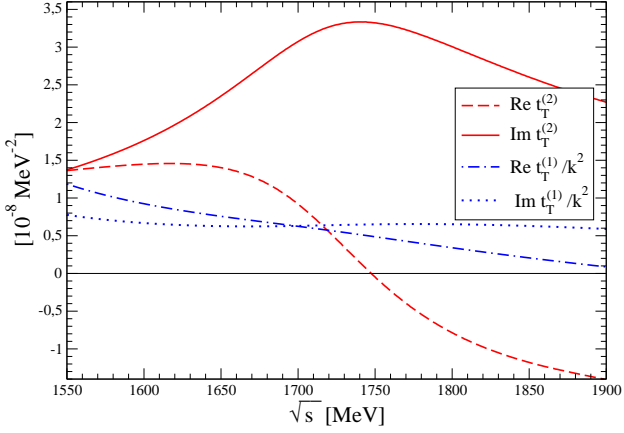}
    \caption{
Functions  $t_T^{(1)}$ and $t_T^{(2)}$   of the triangle diagram.
    }\label{fig:t1t2}
\end{figure}

Since it is the core of the present calculation, it is worth plotting numerically the triangular amplitudes $t_T^{(1)}$ and $t_T^{(2)}$. This is depicted in Fig.~\ref{fig:t1t2}. For $t_T^{(1)}$ we actually plot $t_T^{(1)}/\vec k^2$ to match dimensions with $t_T^{(2)}$.
 We can see a large enhancement in the imaginary part of  $t_T^{(2)}$ at about $1700-1750\mev$ as expected from the previous discussion below Eq.~(\ref{eq:qomenosqa}), stemming from the triangle singularity. The amplitude $t_T^{(1)}$ is not enhanced because the singularity of the loop is produced at the limits of integration of $\cos\theta=-1$ in Eq.~(\ref{eq:ttalpha}) (see ref.~\cite{ncol}). Indeed, the factor $H^{(1)}$ of  
Eq.~(\ref{eq:H1H2}) goes as $1-\cos^2\theta$ which vanishes precisely at $\cos\theta=-1$, canceling the triangle singularity. Therefore, the  enhancement due to the triangle singularity is only manifest in $t_T^{(2)}$.

In the form shown in Eq.~(\ref{eq:tib}), that amplitude cannot be used straightforwardly
as the  $V_{i,\pi\Delta}$ transition potential of  Eq.~(\ref{eq:Vipidelta}) in the coupled channel equation (\ref{eq:T}), since it does not have the 
$\vec \epsilon_i\cdot \vec\epsilon_j$ structure of Eq.~(\ref{eq:Vij}).
However, note that in the coupled channel mechanisms and in order to evaluate a cross section, we would have contributions to a cross section of the kind
\begin{align}
\sum_{M}\left( \vec\epsilon_i\cdot\vec S\, t_T^{(1)}
+ \vec\epsilon_j\cdot \vec k \, \vec S\cdot\vec k  \, t_T^{(2)}
\right)|M\rangle\langle M| \nn\\
\left( \vec\epsilon_j\cdot\vec S^\dagger\, {t_T^{(1)}}^*
+ \vec\epsilon_j\cdot \vec k \, \vec S^\dagger\cdot\vec k  \, {t_T^{(2)}}^*
\right)
\label{eq:sumMa}
\end{align}
with $M$ the third component of the $\Delta$ spin.
Neglecting some spin-flip terms, not of the 
$\vec \epsilon_i\cdot \vec\epsilon_j$ form, Eq.~(\ref{eq:sumMa}) gives

\begin{align}\left(
 \frac{2}{9}\vec k^4 t_T^{(2)}{t_T^{(2)}}^*
+\frac{2}{9}\vec k^2 t_T^{(2)}{t_T^{(1)}}^* \right.\nn\\
\left.+\frac{2}{9}\vec k^2 t_T^{(1)}{t_T^{(2)}}^*
+\frac{2}{3}         t_T^{(1)}{t_T^{(1)}}^*\right)\vec \epsilon_i\cdot \vec\epsilon_j
\label{eq:sumMb}
\end{align}

And then we conclude that $V_\textrm{eff}$ of Eq.~(\ref{eq:Vipidelta})
is such that 
\begin{align}
V_\textrm{eff}^2=  \left(\frac{2}{\sqrt{3}} g \frac{f_{\pi N \Delta}}{m_\pi}
2 M_N\right)^2
\left(
 \frac{2}{9}\vec k^4 t_T^{(2)}{t_T^{(2)}}^* \right.\nn \\
\left.+\frac{2}{9}\vec k^2 t_T^{(2)}{t_T^{(1)}}^*
+\frac{2}{9}\vec k^2 t_T^{(1)}{t_T^{(2)}}^*
+\frac{2}{3}	     t_T^{(1)}{t_T^{(1)}}^*\right)
\label{eq:sumMc}
\end{align}
with $|\vec k|$ the on-shell $\pi$ momentum of the $\pi\Delta$ system with energy $\sqrt{s}$.

So far we have not considered the finite $\Delta$ width which may be relevant because it is large, about 120\mev. This effect can be easily taken into account, on one hand, by folding 
the $\pi\Delta$ loop function  with an analogous expression to  Eq.~(\ref{eq:convo})
but integrating for $\sqrt{s_{\Delta}}$ and using the spectral function 
\be
\rho_{\Delta}(s_{\Delta})=-\frac{1}{\pi}\textrm{Im}
\left\{\frac{1}{\sqrt{s_{\Delta}}-M_{\Delta}+i\frac{\Gamma_{\Delta}}{2}}\right\},
\label{eq:rhoDelta}
\ee
and, on the other hand, by folding Eq.~(\ref{eq:sumMc}) 
by this same $\Delta$ spectral function.

On the other hand, Eq.~(\ref{eq:tib}) and Eq.~(\ref{eq:sumMa}) account both for $\pi\Delta$ S- and D-waves. However, we are also interested in singling out the S-wave contribution to compare the S-wave partial decay width into $\pi\Delta$ with the corresponding column in
Table~\ref{tab:intro}. In order to do this we can trivially rewrite the
$\vec\epsilon\cdot \vec k \, \vec S\cdot\vec k$ 
structure of Eq.~(\ref{eq:tib}) as
\ba
&&\vec\epsilon\cdot \vec k \, \vec S\cdot\vec k= S_i \epsilon_j k_ik_j =\nn\\
&=&S_i \epsilon_j\left[\left(k_ik_j-\frac{1}{3}\vec k^2 \delta_{ij}\right)
+\frac{1}{3}\vec k^2 \delta_{ij}\right].
\ea
Note that the $k_ik_j-\frac{1}{3}\vec k^2 \delta_{ij}$ piece is purely D-wave and 
$\frac{1}{3}\vec k^2 \delta_{ij}$ is purely S-wave. Therefore, if we keep only the S-wave part we arrive to an equivalent expression to  Eq.~(\ref{eq:sumMc}) but which accounts only for the the S-wave $\pi\Delta$ contribution:
\begin{align}
V_\textrm{eff}^2(\textrm{S-wave})=  \left(\frac{2}{\sqrt{3}} g \frac{f_{\pi N \Delta}}{m_\pi}
2 M_N\right)^2
\frac{2}{3}\left|  {t_T^{(1)}}+
 \frac{1}{3}\vec k^2 t_T^{(2)}\right|^2 \,.
\label{eq:sumMswave}
\end{align}

\subsection{The $N(1700)$ couplings to $VB$, $\pi\Delta$, and partial decay widths} 

Usually the couplings of the resonance to the different channels would be obtained from the residues of the amplitudes to the different channels in the second Riemann sheet. However, in the present case the $N(1700)$ is very close to the $\rho N$ threshold and furthermore the $\rho$ width distorts the shape of the amplitude. In such a case it is much better and well defined to obtain the couplings from the amplitudes in the real axis, which actually contain the meaningful information of the resonance. 

As we can see in Fig.~\ref{fig:tii} and we will also see below (see Fig.~\ref{fig:Tij}), the amplitudes resemble pretty much a Breit-Wigner and thus we can parametrize the amplitudes around the resonance mass position as
\be
T_{ij}(\sqrt{s})=\frac{g_i g_j}{\sqrt{s}-M_R+i\frac{\Gamma_R}{2}}\,,
\ee
where $R$ stands for the $N(1700)$ resonance, and $g_i$ is the coupling of the $N(1700)$ to a $VB$ or $\pi\Delta$,
from where the different couplings are related by
\be
g_i=g_j\frac{T_{ij}(M_R)}{T_{jj}(M_R)}
\ee
which allows to obtain all the couplings from one coupling $g_j$ which, up to a global sign, can be obtained from
\be
g_j^2=i\frac{\Gamma_R}{2} T_{jj}(M_R)
\ee
which we choose to be $g_j=g_{\rho N}$, and $M_R$ is the position of the peak of $|T_{jj}|^2$.
From the values of the couplings we can evaluate the partial decay widths of the $N(1700)$ to the different $VB$ and $\pi\Delta$ channels:
\be
\Gamma_i(M_R,m_V)= \frac{1}{2\pi}\frac{M_i}{M_R}|g_i|^2 q
\ee
with $q$ the decay momentum and $M_i$($m_V$) the baryon(vector) mass.
Except for the $\pi\Delta$ channel, the thresholds are above the 
$N(1700)$ mass and thus there is no phase space for the decay to happen
if the resonances were narrow. In those cases the decay takes place thanks to the width of the vector meson and the $N(1700)$, therefore 
we fold the expression
for the width with the mass distribution of the broad particles involved
as 

\begin{align}
\Gamma_{R\to i} = \frac{1}{{\cal N}}
\int_{(M_R-\Gamma_R)^2}^{(M_R+\Gamma_R)^2}ds_R\,
\int_{(M_V-2\Gamma_V)^2}^{(M_V+2\Gamma_V)^2}
ds_V\,\nn\\
\times \rho_R(s_R) \,\rho_V(s_V) \,\Gamma_i(\sqrt{s_R},\sqrt{s_V})
\Theta(\sqrt{s_R}-\sqrt{s_V}-M_i)
\label{eq:convo2}
\end{align}
where $\Theta$ is the step function, with
$q=\frac{1}{2\sqrt{s_R}}\lambda^{1/2}(s_R,s_V,M_i^2)$ 
 and
\begin{align}
{\cal N} =
\int_{(M_R-\Gamma_R)^2}^{(M_R+\Gamma_R)^2}ds_R\,
\int_{(M_V-2\Gamma_V)^2}^{(M_V+2\Gamma_V)^2}
ds_V\,
\rho_R(s_R) \,\rho_V(s_V) .
\label{eq:Nconvo2}
\end{align}
For the $\pi\Delta$ channel the expression is analogous but folding with the $\Delta$ spectral function instead of the vector-meson one.

\section{Results}

\begin{figure}[tbp]
     \centering
     \includegraphics[width=.98\linewidth]{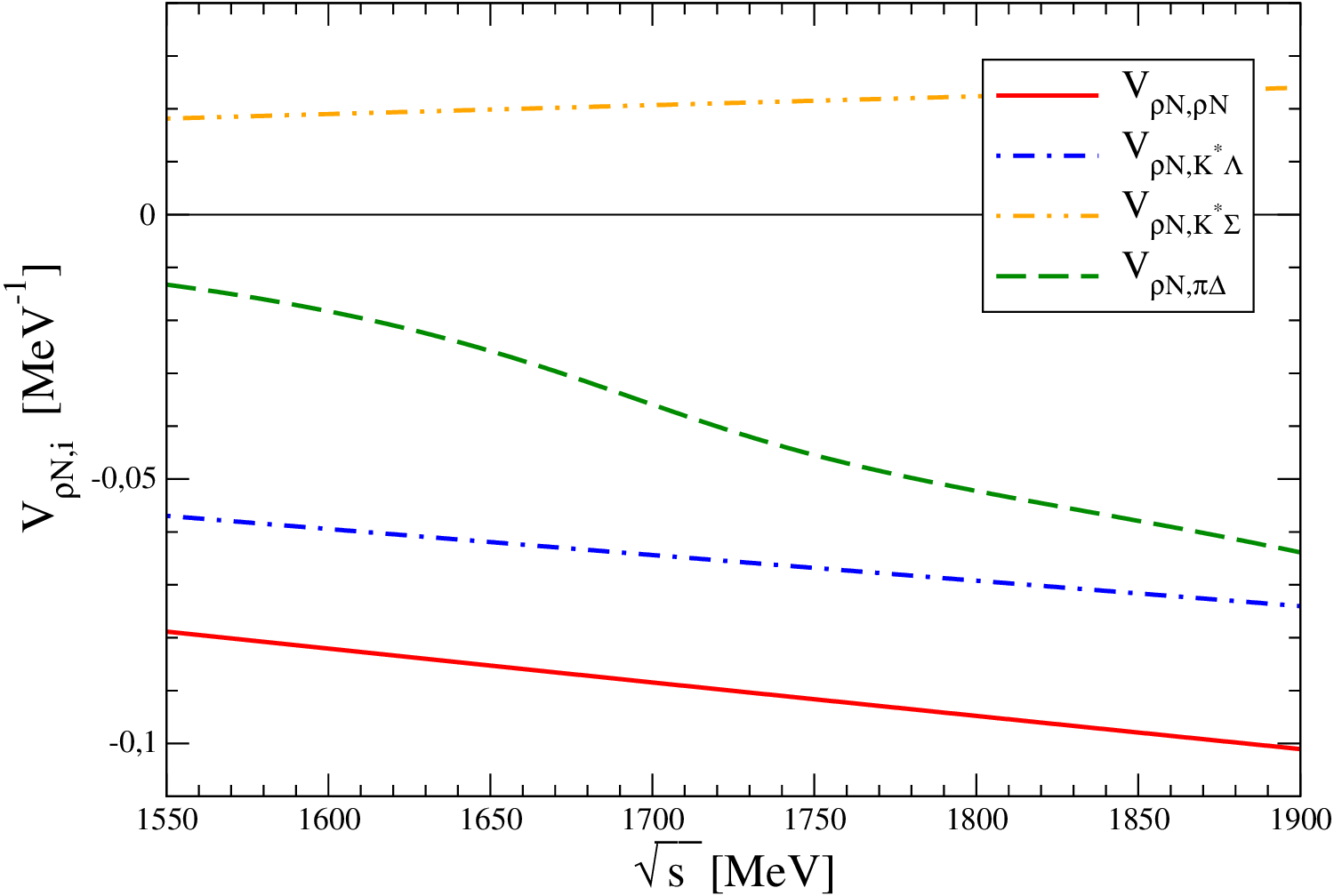}
    \caption{Transition potentials to $\rho N$.}
 \label{fig:Vij}
\end{figure}

First we show in Fig.~\ref{fig:Vij} the value of the transition potentials from the different channels to the dominant one, $\rho N$, in order to foresee the importance of the $\pi\Delta$ channel in the  build up of the $N(1700)$.
The transitions to $\omega N$ and $\phi N$ are zero (see Eq.~(\ref{eq:Cij})).
We see in that figure that the $\rho N\to\pi\Delta$ potential is about 40\% the size of  the $\rho N\to\rho N$ one at the $N(1700)$ peak, which anticipates an important contribution of the $\Delta\pi$ channel in the build up of the resonance.
It is worth noting that the resonance-like shape  from the triangle singularity that we can see in Fig.~\ref{fig:t1t2} is distorted in $V_{\rho N,\pi\Delta}$ due to the
$\vec k^2$ factors in Eq.~(\ref{eq:sumMc}). Nonetheless, it is not the shape of the potential what matters to give the $N(1700)$ resonance shape but its strength, since the resonance comes afterwards from the non-linear coupled channel dynamics.

\begin{figure}[tbp]
     \centering
     \includegraphics[width=.98\linewidth]{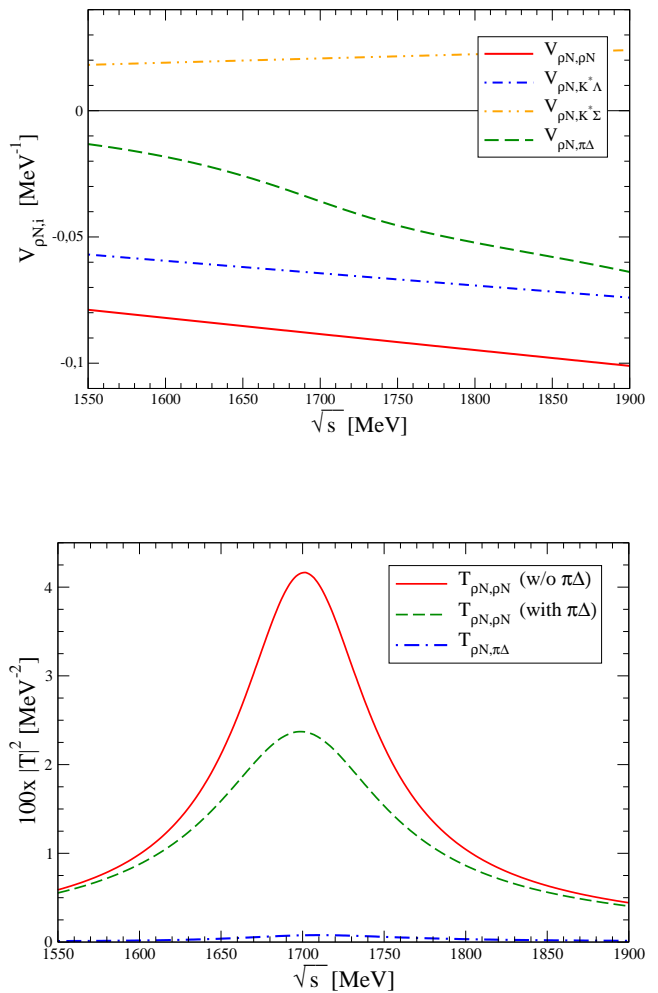}
    \caption{Full amplitudes with and without the $\pi\Delta$ channel.
    }\label{fig:Tij}
\end{figure}

In Fig.~\ref{fig:Tij} we show the modulus squared of $T_{\rho N,\rho N}$, with and without considering the effect of the $\pi\Delta$ channel, and of $T_{\rho N, \pi\Delta}$. For the other channels the effect of considering $\pi\Delta$ is similar but we only show $\rho N$ for clarity and because $\rho N$ is the dominant one.
 We can see that, although the $T_{\rho N, \pi\Delta}$ amplitude is small, the effect of the $\pi\Delta$ channel in the highly non-linear dynamics involved in the coupled channels produces a significant broadening of the $N(1700)$ resonance. Actually, by fitting to a Breit-Wigner shape, the mass and width of the $N(1700)$ resonance before considering the $\pi\Delta$ channel was $M_R=1701\mev$, $\Gamma_R=105\mev$, and after including the $\pi\Delta$ channel they are
$M_R=1699\mev$, $\Gamma_R=141\mev$.
While the mass barely changes, the width increases about 35\% from the inclusion of the $\pi\Delta$ channel. As can be seen in Table~\ref{tab:intro}, the width of the $N(1700)$ resonance in the PDG \cite{pdg} is quite uncertain $100-250\mev$, and it seems that a large width is preferred, for which the inclusion of the $\pi\Delta$ channel goes in the right direction.

In Table~\ref{tab:coupwidths} we show the couplings and partial decay widths, with and without including the $\pi\Delta$ channel.

\begin{table*}[ht]
\begin{center}
\begin{tabular}{|c|c|c|c|c|c|c|c|}
\hline 
 & \multicolumn{3}{c|}{
 \begin{tabular}[c]{c} without $\pi\Delta$\\  
 $M_R=1701$ \\ $\Gamma_R=104$ 
  \end{tabular} }
 &\multicolumn{4}{c|}{
 \begin{tabular}[c]{c} with $\pi\Delta$\\  
 $M_R=1698$\\ $\Gamma_R=141$  
  \end{tabular} }  \\
\cline{2-8}
& $g_i$&  $|g_i|$ & $\Gamma_i$  &  $g_i$ &  $|g_i|$ & $\Gamma_i$  & $\Gamma_i/\Gamma_R$ \\
\hline  &&&& \\[-4.5mm]
$\rho N$        &   $3.07-1.09i $   & 3.26  & 108 &   $2.98-1.35i $  & 3.27	 & 114  & 80\%   \\ \hline
$\omega N$      &   $0.17-0.06i $   & 0.18  & 0.2 &   $0.17-0.08i $  & 0.19      & 0.3  &  0.2\% \\ \hline
$\phi N$        &   $-0.26+0.09i$   & 0.28 & 0	  &   $-0.26+0.12i$  & 0.28	 & 0    &  0 \\  \hline
$K^* \Lambda$   &   $2.61-0.92i $   & 2.77  & 0   &   $2.55-1.15i $  & 2.80      & 0.1  &  0.1\% \\ \hline
$K^* \Sigma$    &   $-0.63+0.22i$   & 0.67 & 0	  &   $-0.61+0.28i$  & 0.67	 & 0    &  0 \\ \hline
$\pi\Delta$     & --                &   --  & --  &   $0.16+0.57i$   & 0.61      & 14.6 &  10\% \\ \hline
 \begin{tabular}[c]{c}$\pi\Delta$\\  (S-wave) \end{tabular} 
                & --                & --   & --   &   $-0.28 +i0.26$ & 0.40      & 6.4  &  5\% \\ \hline
\end{tabular}
\caption{Couplings and partial decay widths of the $N(1700)$ resonance. All dimensional magnitudes are in MeV.}
\label{tab:coupwidths}
\end{center}
\end{table*}

The decay width to $\pi\Delta$ is relevant despite the small value of its coupling because of the larger phase space available for this channel.
Note that the partial decay widths do not sum up exactly to the total width of the $N(1700)$ resonance since the convolution we are carrying out
to get the $VB$ partial decay widths is just an approximation and the resonance is very close to the $\rho N$ threshold.
However, the sum of partial decay widths, 129~MeV, is close to the total width of 141~MeV.
 This discrepancy should be understood as a measure of the uncertainty in our calculation of the decay widths.
 
It is worth mentioning that the analyses \cite{sokho,ani} that get a large  branching ratio to $\pi\Delta$ in Table~\ref{tab:intro}, fit data with $\pi\pi N$ in the final state but only for two neutral $\pi^0$, and the $\rho$ meson does not couple to $\pi^0\pi^0$. Therefore they do not have the $\rho N$ channel in their analyses and thus it is easy to overestimate the weight of the $\pi\Delta$. However in the analysis of ref.~\cite{shre}, $\pi\pi N$ data with charged pions
 are also considered and thus the $\rho N$ weight is also obtained. When the $\rho N$ channel is considered, the importance of the $\pi\Delta$ is reduced considerably in the direction of what we get in the present work. Furthermore note that, as already mentioned in the introduction, the mass obtained in refs.~\cite{sokho,ani} for the $N(1700)$ resonance lays around 1800$\mev$ and the width is also much larger than in the other analyses. This could be an indication that the state around 1700 MeV and the one around 1800 MeV are actually two different resonances and thus the coupling to $\pi\Delta$ could be very different.

In this context it is interesting to note that the analysis of Ref.~\cite{shre}, which gives a sizeable fraction of $\rho N$ decay, $38\pm 6 \%$, also provides a small mass for the resonance, and one should accept that  this channel, being only allowed because of the width of the $\rho$ meson, is not particularly easy to determine quantitatively, which means that part of its strength can be easily attributed to other channels.  The results of our study clearly indicate a large dominance of $\rho N$ in the build up of the resonance and also in its decay width. However, it is also clear that the $\pi\Delta$ channel
has a significant contribution which theoretically we can
only explain from the triangular singularity studied in
the present work.  These findings, and the large dispersion of the experimental results concerning the $N(1700)$, should serve to stimulate further experimental work to the light of the present results. This should also include some modification in the analysis tools to explicitly include the structure tied to the triangle singularity, which, as we have discussed here, is tied to the masses of the particles in the triangle diagram and, hence, is unavoidable.

\section{Summary and conclusions}

In the present work we have shown how a channel, which otherwise would be negligible, can gain unsuspected importance in the build up of a resonance thanks to an  enhancement in a triangular loop from a mathematical singularity appearing for particular kinematic conditions. Specifically we show that the $\pi\Delta$ channel has an important relevance in the $N(1700)$ formation of the resonance because of a singularity, of non-resonant nature, in the Feynman diagram of Fig.~\ref{fig:triang}. We have shown how the $N(1700)$ is dynamically generated from the implementation of unitarity from the interaction of the octet of vector mesons and the octet of baryons in coupled channels and how the $VB$ can couple effectively to $\pi\Delta$ through the triangle diagram with a $\rho$ meson, a nucleon and a pion in the internal lines. For the values of the masses of the particles involved, the loop has a singularity at about the same energy where the $N(1700)$ peaks.  This makes the $\pi\Delta$ channel gain an importance which otherwise would be negligible. 
 We obtain an increase of about 35\% in the width of the $N(1700)$ resonance when the $\pi\Delta$ channel is taken into account and make predictions for the partial decay widths into the different $VB$ channels and the $\pi\Delta$ channel, considering also the S-wave separately in this latter channel.
 
 We stressed the fact that presently there is a large dispersion in the data of the different analyses, and, hence, a quantitative comparison with our results would be inconclusive.  Instead, we pointed out at the convenience of a reanalysis of the data to the light of the findings of the present work, which include the unavoidable presence of a triangle singularity. To account for it, a modification of the tools normally used in experimental analyses would have to be implemented,  to include the structure of the triangle mechanism as an extra multiplicative factor in the $\pi\Delta$ channel. The continuous accumulation of data in the $\pi\pi N$ channel, and improved analysis tools, should bring more precise information on this resonance in the future, allowing us to learn more about its nature.

\section*{Acknowledgments}

This work is partly supported by the Spanish Ministerio
de Economia y Competitividad and European FEDER funds
under the contract number FIS2011-28853-C02-01, FIS2011-
28853-C02-02, FIS2014-57026-REDT, FIS2014-51948-C2-
1-P, and FIS2014-51948-C2-2-P, and the Generalitat Valenciana
in the program Prometeo II-2014/068 (EO).

\end{document}